%% file: main.tex
\documentclass[sigconf, 9pt]{acmart}

\usepackage{microtype}
\usepackage{graphicx}
\usepackage{booktabs}
\usepackage{hyperref}
\usepackage{array}
\usepackage{multirow}
\usepackage{algorithm}
\usepackage{algorithmic}
\usepackage{subcaption}
\usepackage{amsfonts}

\newcommand{\red}[1]{\textcolor[rgb]{1.0, 0.1, 0.5}{\textbf{#1}}}

\AtBeginDocument{%
  \providecommand\BibTeX{{%
    \normalfont B\kern-0.5em{\scshape i\kern-0.25em b}\kern-0.8em\TeX}}}

\setcopyright{acmcopyright}
\copyrightyear{2022}
\acmYear{2022}
\acmDOI{XXXXXXX.XXXXXXX}

\acmConference[ICCAD 2022]{International Conference on Computer-Aided Design}{Oct. 30--Nov. 3,
  2022}{San Diego, CA}
\acmPrice{XX.XX}
\acmISBN{XXX-X-XXXX-XXXX-X/XX/XX}

\begin{document}

\title[Computing-In-Memory Neural Network Accelerators for Safety-Critical Systems: Can Small Device Variations Be Disastrous?]{\huge{Computing-In-Memory Neural Network Accelerators for Safety-Critical Systems: Can Small Device Variations Be Disastrous?}}

\author{Zheyu Yan} 
\email{zyan2@nd.edu}
\affiliation{%
  \institution{University of Notre Dame}
  \country{}
}
\author{Xiaobo Sharon Hu}
\email{shu@nd.edu}
\affiliation{%
  \institution{University of Notre Dame}
  \country{}
}
\author{Yiyu Shi}
\email{yshi4@nd.edu}
\affiliation{%
  \institution{University of Notre Dame}
  \country{}
}


\begin{abstract} 
Computing-in-Memory (CiM) architectures based on emerging non-volatile memory (NVM) devices have demonstrated great potential for deep neural network (DNN) acceleration thanks to their high energy efficiency. However, NVM devices suffer from various non-idealities, especially device-to-device variations due to fabrication defects and cycle-to-cycle variations due to the stochastic behavior of devices. As such, the DNN weights actually mapped to NVM devices could deviate significantly from the expected values, leading to large performance degradation. To address this issue, most existing works focus on maximizing average performance under device variations. This objective would work well for general-purpose  scenarios. But for safety-critical applications, the worst-case performance must also be considered.  Unfortunately, this has been rarely explored in the literature. In this work, we formulate the 
problem of determining the worst-case performance of CiM DNN accelerators under the impact of device variations. We further propose a method to effectively find the specific combination of device variation in the high-dimensional space that leads to the worst-case performance. We find that even with very small device variations, the accuracy of a DNN can drop drastically, causing concerns when deploying CiM accelerators in safety-critical applications. Finally, we show that surprisingly none of the existing methods used to enhance average DNN performance in CiM accelerators are very effective when extended to enhance the worst-case performance, and further research down the road is needed to address this problem.
\end{abstract}

\maketitle

\input{M1_Introductions}
\input{M2_Preliminaries}

\input{M3_GPO}
\input{M4_GPO_Experiments}
\input{M5_Protect}
\input{M6_Conclusions}

\clearpage
\bibliographystyle{ACM-Reference-Format}
\bibliography{M7_References}

\end{document}

%% file: M1_Introductions.tex
\section{Introductions}

Deep Neural Networks (DNNs) have reached superhuman performance in a variety of perception tasks including speech recognition, object detection, and image classification~\cite{liang2022variational, liang2021can, wang2021exploration}. Thus, there is an obvious trend to use DNNs to empower edge applications in smart sensors, smartphones, automobiles, and \emph{etc.}~\cite{yang2020co, lu2022semi, sheng2022larger} However, because of the constrained computation resources and limited power budget of edge platforms, CPUs or GPUs are not always good candidate computing units for implementing computation-intensive DNNs on edge devices.

Computing-in-Memory (CiM) DNN accelerators~\cite{shafiee2016isaac} is a great alternative candidate for edge DNN implementation because they can reduce data movement by performing in-situ weight data access~\cite{sze2017efficient}. Furthermore, by employing non-volatile memory (NVM) devices (\emph{e.g.}, ferroelectric field-effect transistors (FeFETs), resistive random-access memories (RRAMs), and phase-change memories (PCMs)), CiM can achieve higher memory density and higher energy efficiency compared with conventional MOSFET-based designs~\cite{chen2016eyeriss}.
However, NVM devices can be unreliable, especially because of device-to-device variations due to fabrication defects and cycle-to-cycle variations due to the stochastic behavior of devices. If not handled properly, the weight values provided by the NVM devices during computations could deviate significantly from the expected values, leading to great performance degradation. 

To quantify the robustness of CiM DNN accelerators, a Monte Carol (MC) simulation-based evaluation process is often adopted~\cite{peng2019dnn+}. A device variation model is extracted from physical measurements. Then in each MC run, one instance of each device is sampled from the variation model and DNN performance evaluation is collected. This process is repeated thousands of times until the collected DNN performance distribution converges. Following this process, existing practices~\cite{yan2020single, jin2020improving, liu2019fault, he2019noise, yan2022swim} generally include up to 10,000 MC runs, which is extremely time-consuming. Other researchers use Bayesian Neural Networks (BNNs) to evaluate the robustness against device variations~\cite{gao2021bayesian}, but the variational inference of BNNs is essentially one form of MC simulation.  

Based on these evaluation methods, many works have been proposed in the literature to improve the average performance of CiM DNN accelerators under device variations. They fall into two categories: (1) reducing device variations and (2) enhancing DNN robustness. To reduce device variations, a popular option is write-verify~\cite{shim2020two}. The approach applies iterative write and read (verify) pulses to make sure that the maximum difference between the weights eventually programmed into the devices and the desired values are bounded by a designer-specified threshold. Write-verify can reduce the weight deviation from the ideal value to less than 3\%, thus reducing the average accuracy degradation of deployed DNNs to less than 0.5\%~\cite{shim2020two}. To enhance DNN robustness, a variety of approaches exist. For example, neural architecture search is devised~\cite{yan2021uncertainty, yan2022radars} to automatically search through a designated search space for DNN architectures that are more robust. Variation-aware training~\cite{jiang2020device, he2019noise, chang2019nv}, on the other hand, injects device variation-induced weight perturbations in the training process, so that the trained DNN weights are more robust against similar types of variations. Other approaches include on-chip in-situ training~\cite{yao2020fully} that trains DNNs directly on noisy devices and Bayesian Neural Network (BNN)-based approaches that utilizes the variational training process of BNNs to improve DNN robustness~\cite{gao2021bayesian}.\footnote{This project is supported in part by NSF under grants CNS-1822099, CCF-1919167 and CCF-2028879.}

\begin{figure}[ht]
    \includegraphics[trim=0 150 130 0, clip, width=0.95\linewidth]{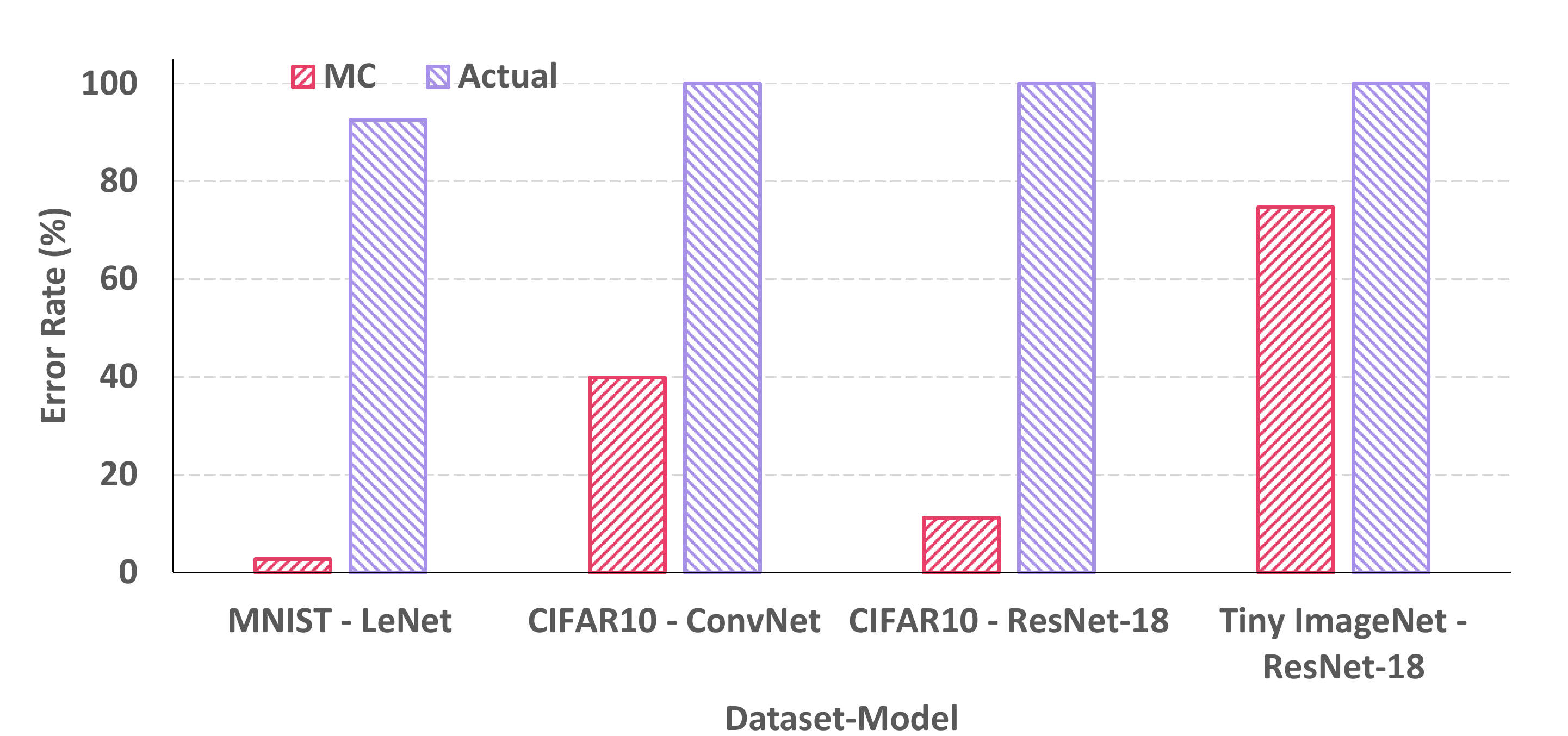}
    \vspace{-0.4cm}
    \caption{Comparison of the worst-case (highest) top-1 error identified by 100K MC runs and our method, based on device variation model introduced in Section~\ref{sect:model}. Showing results on four DNNs: LeNet for MNIST, CovNet for CIFAR-10, ResNet-18 for CIFAR-10, and ResNet-18 for Tiny ImageNet. }
    \label{fig:Intro}
    \vspace{-0.4cm}
\end{figure}

One common assumption of all these evaluation and improvement methods is that they focus on the average performance of a DNN under device variations, which may work well in general-purpose scenarios. However, for safety-critical applications where failure could result in loss of life (\emph{e.g.}, medical devices, aircraft flight control, and nuclear systems), significant property damage, or damage to the environment, only focusing on the average performance is not enough. The worst-case performance, regardless of its likelihood, must also be considered~\cite{wang2022efficient, juba2020more}. Yet this is a very challenging problem:  
Given the extremely high dimension of the device variation space, simply running MC simulations in hope to capture the worst-case corner during training will not work. As shown in Fig.~\ref{fig:Intro}, for various DNNs for different datasets, even though the MC simulation has converged with 100K runs, the highest DNN top-1 error rate discovered is still much higher than that identified by our method to be discussed later in this paper, where the worst-case error is close to 100\%. 


Despite the importance of the problem, very little has been explored in the literature. The only related work comes from the security perspective~\cite{tsai2021formalizing}, where a weight projected gradient descent attack (PGD) method is developed to find the weight perturbation that can lead to mis-classification of inputs. However, the goal is to generate a successful weight perturbation attack, but not to identify the worst-case scenario under all possible variations.


To fill the gap, in this work we propose an optimization framework that can efficiently and effectively find the worst-case performance of DNNs in a CiM accelerator with maximum weight variations bounded by write-verify. We show that the problem can be formulated as a constrained optimization with non-differentiable objective, which can be relaxed and solved by gradient-based optimization. We then conduct experiments on different networks and datasets under a practical setting commonly used (\emph{i.e.}, each device represents 2 bits of data with write-verify and yields a maximum weight perturbation magnitude of 0.03). Weight PGD \cite{tsai2021formalizing}, the only method of relevance in the literature, identifies worst-case scenarios where the accuracy is similar to that of random guess, while ours can find ones with close to zero accuracy. We then use our framework to evaluate the effectiveness of existing solutions designed to enhance the average accuracy of DNNs under device variations, and see how they improve the worst-case performance. We study two types of remedies: reducing device variations and enhancing DNN robustness. Experimental results suggest that they either induce significant overhead or are not quite effective, and further research is needed. 

The main contributions of this work are multi-fold: 
\begin{itemize}
    \item This is the first work that formulates the problem of finding worst-case performance in DNN CiM accelerators with device variations for safety-critical applications. 
    \item An efficient gradient-based framework is proposed to solve the non-differentiable optimization problem and find the worst-case scenario.  
    \item Experimental results show that our framework is the only method that can effectively identify the worst-case performance of a DNN.   
    \item We show that even though the maximum weight perturbations are bounded (\emph{e.g.}, by write-verify) to be very small, significant DNN accuracy drop can still occur. Therefore any application of CiM accelerators in the safety-critical settings should use caution.
    \item We further demonstrate that existing methods to enhance DNN robustness are either too costly or not quite effective in improving worst-case performance. New algorithms in this direction are needed. 
\end{itemize}

The remainder of the paper is organized as follows. In Section~\ref{sect:related} we first discuss the background information about CiM DNN accelerators, their robustness issue caused by device variations, and existing methods targeting this issue. We then formulate the problem of finding the worst-case performance of DNN under device variations and propose a framework to solve it in Section~\ref{sect:proposed}, along with experimental results to show its efficacy. Extensive studies on the effectiveness of extending existing methods to enhance DNN worst-case performance are carried out in Section~\ref{sect:protect} and concluding remarks are given in Section~\ref{sect:conclusion}. 

%% file: M2_Preliminaries.tex
\section{Related Works}\label{sect:related}

\subsection{Crossbar-based Computing Engine}\label{sec:2.1}
\begin{figure}[ht]
\vspace{-0.3cm}
\begin{center}
\centerline{\includegraphics[trim=0 150 550 0, clip, width=0.4\linewidth] {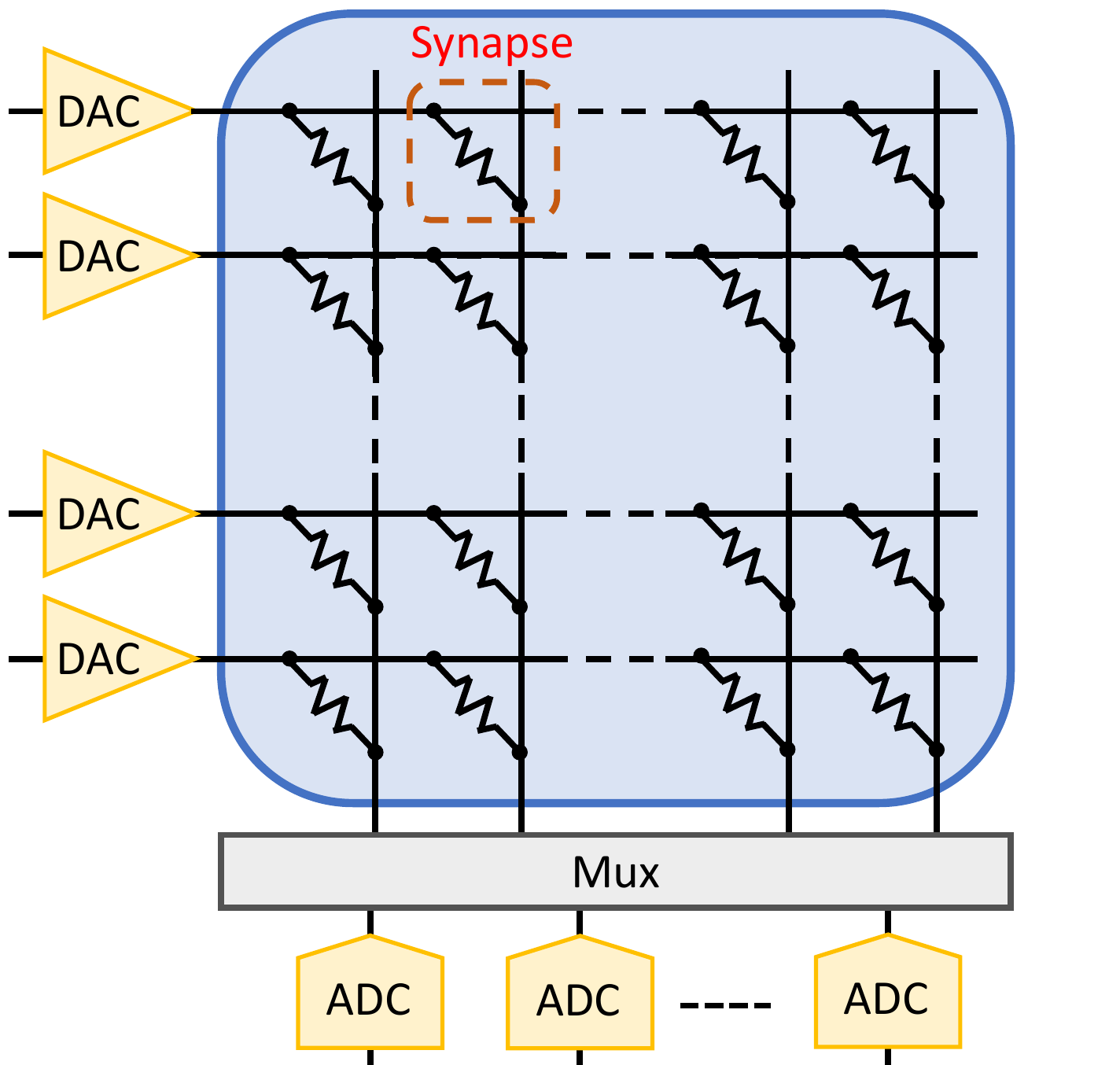}}
\vspace{-0.4cm}
\caption{Illustration of the crossbar architecture. The input is fed horizontally and multiplied by weights stored in the NVM devices at each cross point. The multiplication results are summed up vertically and the sum serves as an output.}
\vspace{-0.5cm}
\label{fig:crossbar}
\end{center}
\end{figure}

The crossbar array is the key computation engine of CiM DNN accelerators. A crossbar array can perform matrix-vector multiplication in one clock cycle. In a crossbar array, matrix values (\emph{e.g.}, weights in DNNs) are stored at the cross point of each vertical and horizontal line with NVM devices (\emph{e.g.}, RRAMs and FeFETs), and each vector value (\emph{e.g.}, inputs for DNNs) is fed in through horizontal data lines in the form of voltage. The output then flows out through vertical lines in the form of current. The calculation in the crossbar array is performed in the analog domain according to the Kirchhoff's laws but additional peripheral digital circuits are needed for other key DNN operations (\emph{e.g.}, pooling and non-linear activation), so digital-to-analog and analog-to-digital converters are used between different components especially for DACs to transfer digital input data to voltage levels and ADCs to transfer analog output currents into digital values.

Resistive crossbar arrays suffer from various sources of variations and noises. Two major ones include spatial variations and temporal variations. Spatial variations result from fabrication defects and have both local and global correlations. NVM devices also suffer from temporal variations due to the stochasticity in the device material, which causes fluctuations in conductance when programmed at different times. Temporal variations are typically independent of the device to device and are irrelevant to the value to be programmed~\cite{feinberg2018making}.  
In this work, as a proof of concept, we assume the impact of these non-idealities to be independent on each NVM device. The proposed framework can also be extended to other sources of variations with modification.

\subsection{Evaluating DNN Robustness Against Device Variations}

Most existing researches use a Monte Carlo (MC) simulation-based evaluation process to quantify the robustness of CiM DNN accelerators under the impact of device variations. A device variation model and a circuit model are first extracted from physical measurements. The DNN to be evaluated is then mapped onto the circuit model and the desired value of each NVM device is calculated. In each MC run, for each device, one instance of a non-ideal device is sampled from the variation model, and the actual conductance value of each NVM device is determined. After that, DNN performance (\emph{e.g.}, classification accuracy) can be collected. This process is repeated thousands of times until the collected DNN performance distribution converges. Existing practices~\cite{yan2020single, yan2021uncertainty, he2019noise, liu2019fault, jin2020improving, yan2022swim} generally include close to 10,000 MC runs, which is extremely time-consuming. Empirical results~\cite{yan2020single, yan2021uncertainty} show that 10k MC runs are enough for evaluating the average accuracy of DNNs while no theoretic guarantee is provided. 

Several researchers have also looked into the impact of weight perturbations on neural network security~\cite{wu2020adversarial, tsai2021formalizing}. This line of research, dubbed as ``Adversarial Weight Perturbation'', tries to link the perturbation in weight to the more thoroughly studied adversarial example issue, where the inputs of DNNs are intentionally perturbed to trigger mis-classifications. One work~\cite{wu2020adversarial} trains DNNs on adversarial examples to collect adversarial weight perturbation. Most recently, \cite{tsai2021formalizing} tries to find the adversarial weight perturbation through a modified weight Projected Gradient Descent (PGD) attack. The method can successfully find a small perturbation to reduce the accuracy of DNNs. 
The work focuses on the success of the attack and does not offer a guarantee of worst-case weight perturbation, as will be demonstrated by our experimental results. 

\subsection{Addressing Device Variations}\label{sec:2.3}
Various approaches have been proposed to deal with the issue of device variations in CiM DNN accelerators. Here we briefly review the two most common types: enhancing DNN robustness and reducing device variations. 

A common method used to enhance DNN robustness against device variations is variation-aware training\cite{jiang2020device,peng2019dnn+,he2019noise,chang2019nv}. Also known as noise-injection training, the method injects variation to DNN weights in the training process, which can provide a DNN model that is statistically robust against the impact of device variations. In each iteration, in addition to traditional gradient descent, an instance of variation is sampled from a variation distribution and added to the weights in the forward pass. The backpropagation pass is noise free. Once the gradients are collected, this variation is cleared and the variation-free weight is updated according to the previously collected gradients. Other approaches include designing more robust DNN architectures~\cite{jiang2020device, yan2021uncertainty, gao2021bayesian, ye2021bayesft} and pruning~\cite{jin2020improving, chen2021pruning}.

To reduce device variations, write-verify\cite{shim2020two, yao2020fully} is commonly used
during the programming process. An NVM device is first programmed to an initial state using a pre-defined pulse pattern. Then the value of the device is read out to verify if its conductance falls within a certain margin from the desired value (\emph{i.e.}, if its value is precise). If not, an additional update pulse is applied, aiming to bring the device conductance closer to the desired one. This process is repeated until the difference between the value programmed into the device and the desired value is acceptable. The process typically requires a few iterations. Most recently, researchers have demonstrated that it is possible to only selectively write-verify a small number of critical devices to maintain the average accuracy of a DNN \cite{yan2022swim}. There are also various circuit design efforts~\cite{shin2021fault, li202140nm, jeong2022variation} that try to 
mitigate the device variations.

%% file: M3_GPO.tex
\section{Evaluating Worst-Case Performance of CiM DNN Accelerators}\label{sect:proposed}
A major impact of the device variations is that the conductance of the NVM devices will deviate from the desired value due to the device-to-device variations and cycle-to-cycle variations during the programming process, leading to perturbations in the weight values of a DNN and affecting its accuracy. 
In Section~\ref{sect:model}, we first model the impact of NVM device variations on weight perturbation, assuming write-verify is used to minimize the variations. Then based on the weight perturbation model, in Section~\ref{sect:problem}, we formulate the problem of finding the lowest DNN accuracy under weight perturbation, and devise a framework to solve it. Experimental results are presented in Section~\ref{sect:method}. 

\subsection{Modeling of Weight Perturbation Due to Device Variations}\label{sect:model}
Here we show how we model the impact of device variations on DNN weights. In this paper, we are majorly concerned about the impact of device variations in the programming process, \emph{i.e.}, the conductance value programmed to NVM devices is different from the desired value.

For a weight represented by $M$ bits, let its desired value $\mathcal{W}_{des}$ be
\begin{equation}
    \mathcal{W}_{des} = \sum_{j=0}^{M-1}{m_j \times 2^j}
\end{equation}
where $m_j$ is the value of the $j^{th}$ bit of the desired weight value. Moreover, each NVM device is capable of representing $K$ bits of data. Thus, each weight value of the DNNs needs $M/K$ devices to represent\footnote{Without loss of generality, we assume that M is a multiple of K.}. This mapping process can be represented as
\begin{equation}
    g_i = \sum_{j=0}^{K -1} m_{i\times K + j} \times 2^j
\end{equation}
where $g_i$ is the desired conductance of the $i^{th}$ device representing a weight. Note that negative weights are mapped in a similar manner.
We assume all of the devices use write-verify so that the difference between the actual conductance of each device and its desired value is bounded~\cite{feinberg2018making}:
\begin{equation}
    gp_i = g_i + n_i, -th\leq n_i \leq th
\end{equation}
where $gp_i$ is the actually programmed conductance and $th$ is the designated write-verify tolerate threshold.

Thus, when a weight is programmed, the actual value $\mathcal{W}_{p}$ mapped on the devices would be

\begin{align}
\begin{split}
    \mathcal{W}_{p}     & = \sum_{i=0}^{M/K -1}2^{i\times K}{gp_i } \\
                        & = \sum_{i=0}^{M/K -1}2^{i\times K}{g_i + n_i} \\
                        & = \mathcal{W}_{des} + \sum_{i=0}^{M/K-1}{n_i \times 2^{i\times K}}\\
    \mathcal{W}_{des} - th_g &\leq \mathcal{W}_{p} \leq \mathcal{W}_{des} + th_g\label{eq:noise} 
\end{split}
\vspace{0.0 cm}
\end{align}
where $th_g = \sum_{i=0}^{M/K-1}{\left(th \times 2^{i\times K}\right)}$. In this paper, we denote $th_g$ as the {\em weight perturbation bound}.

In this paper, we set $K=2$ as in~\cite{jiang2020device, yan2022swim}. Same as the standard practice discussed in Section~\ref{sec:2.3}, for each weight, we iteratively program the difference between the value on the device and the expected value until it is below 0.1, \emph{i.e.}, $th = 0.06$~\cite{yan2022swim}, resulting in $th_g = 0.03$ (unless otherwise specified). These numbers are in line with those reported in~\cite{shim2020two}, which confirms the validity of our model and parameters.

\subsection{Problem Definition}\label{sect:problem}

Now that we have the weight perturbation model, we can define the problem of identifying the worst-case DNN accuracy. Without loss of generality, in this work, we use $\{f, \mathbf{W}\}$ to represent a neural network, where $f$ is its neural architecture and $\mathbf{W}$ is its weights. The forward pass of this neural network is represented by $f(\mathbf{W},\mathbf{x})$, where $\mathbf{x}$ is the inputs.

From the model in Section~\ref{sect:model}, we can see that the weight perturbation due to device variations is additive and independent. As such, the forward pass of a neural network under the impact of device variation can be expressed as $f(\mathbf{W} + \Delta\mathbf{W},\mathbf{x})$, where $\Delta\mathbf{W}$ is the weight perturbation caused by device variations. We can define the \textit{perturbed neural network} as $\{f, \mathbf{W} + \Delta\mathbf{W}\}$.

With these annotations, we can have the following problem definition: Given a neural network $\{f, \mathbf{W}\}$ and an evaluation dataset $D$, find the perturbation $\Delta\mathbf{W}$ that the accuracy of perturbed neural network $\{f, \mathbf{W} + \Delta\mathbf{W}\}$ in dataset $D$ is the lowest among all possible perturbations inside the weight perturbation bound. In the rest of this paper, this perturbation $\Delta\mathbf{W}$ is denoted as the \textit{worst-case weight perturbation}; the resultant performance(accuracy) is denoted as the \textit{worst-case performance(accuracy)}; and the corresponding neural network is denoted as the \textit{worst-case neural network}. 

Under this definition, we can formalize the problem as:
\begin{equation}
\begin{split}
    \underset{\Delta\mathbf{W}}{\mathrm{minimize}} & \ \ \ \  |\{f(\mathbf{W}+\Delta\mathbf{W},\mathbf{x}) == t | (x,t)\in D\}| \\
    \mathrm{s.t.}     & \ \ \ \  \mathcal{L}(\Delta\mathbf{W}) \leq th_g\label{eq:ori_def}
\end{split}
\end{equation}
where $x$ and $t$ are the input data and classification label in dataset $D$, respectively. $\mathcal{L}(\Delta\mathbf{W})$ is the maximum magnitude of weight perturbation, \emph{i.e.}, $\max(abs(\Delta\mathbf{W}))$, $th_g$ is the weight perturbation bound in (\ref{eq:noise}) in Section~\ref{sect:model}, and $|A|$ denotes the cardinality (size) of a set $A$. As $f$, $\mathbf{W}$ and $D$ are fixed, the goal is to find the $\Delta\mathbf{W}$ that minimizes the size of this set of correct classifications, \emph{i.e.}, achieving the worst-case accuracy.

\subsection{Finding the Worst-Case Performance}\label{sect:method}

The optimization problem defined by (\ref{eq:ori_def}) is extremely difficult to solve directly due to the non-differentiable objective function. In this section, we put forward a framework to cast it into an alternative form that can be solved by existing optimization algorithms.

To start with, we can slightly relax the objective. Consider a function $p$ such that $f(\mathbf{W}+\Delta\mathbf{W},\mathbf{x}) == t$ if and only if $p(\mathbf{x}, \{f, \mathbf{W}+\Delta\mathbf{W}\}) \geq 0$. In this case, the optimization objective
\begin{equation}
    |{f(\mathbf{W}+\Delta\mathbf{W},\mathbf{x}) == t | (x,t)\in D\}}| \label{eq:orig}
\end{equation}
can be relaxed to
\begin{equation}
    \sum_{\mathbf{x}\in D} p(\mathbf{x}, \{f, \mathbf{W}+\Delta\mathbf{W}\}) \label{eq:relaxed}
\end{equation}
Intuitively, minimizing (\ref{eq:relaxed}) can help to minimize (\ref{eq:orig}), 
and these two optimization problems become strictly equivalent if all data in $D$ is mis-classified in the presence of $\Delta\mathbf{W}$. 

There are various choices of $p(\mathbf{x}, \{f, \mathbf{W}+\Delta\mathbf{W}\})$ that can meet the requirement. We show some of the representative ones below. 
\begin{equation}
\begin{split}
    \mathbf{O} = & \ f(\mathbf{W}+\Delta\mathbf{W},\mathbf{x}) \\
    \mathbf{Z} = & \ \mathrm{Softmax}(f(\mathbf{W}+\Delta\mathbf{W},\mathbf{x})) \\
    p_1(\mathbf{x}, \{f, \mathbf{W}+\Delta\mathbf{W}\}) = & -loss(\mathbf{O}, t) + 1\\
    p_2(\mathbf{x}, \{f, \mathbf{W}+\Delta\mathbf{W}\}) = & \max\{\max_{i\neq t}(O_i) - O_{t}, 0\}\\
    p_3(\mathbf{x}, \{f, \mathbf{W}+\Delta\mathbf{W}\}) = & \ \mathrm{softplus}(\max_{i\neq t}(O_i)-O_t) - \log(2)\\
    p_4(\mathbf{x}, \{f, \mathbf{W}+\Delta\mathbf{W}\}) = & \max\{0.5-O_t,0\}\\
    p_5(\mathbf{x}, \{f, \mathbf{W}+\Delta\mathbf{W}\}) = & -\log(2 \cdot O_t -2)\\
    p_6(\mathbf{x}, \{f, \mathbf{W}+\Delta\mathbf{W}\}) = & \max\{\max_{i\neq t}(Z_i) - Z_{t}, 0\}\\
    p_7(\mathbf{x}, \{f, \mathbf{W}+\Delta\mathbf{W}\}) = & \ \mathrm{softplus}(\max_{i\neq t}(Z_i)-Z_t) - \log(2)\\
\end{split}
\end{equation}
where $x$ and $t$ are the input data and classification label, respectively. $\mathrm{softplus}(x) = \log(1+\exp(x))$ and $loss(\mathbf{O}, t)$ is cross entropy loss.

According to the empirical results, in this paper we choose
\begin{equation}
\begin{split}
    \mathbf{O} = & \ f(\mathbf{W}+\Delta\mathbf{W},\mathbf{x}) \\
    p(\mathbf{x}, \{f, \mathbf{W}+\Delta\mathbf{W}\}) = & \max\{\max_{i\neq t}(O_i) - O_{t}, 0\}
\end{split}
\end{equation}

We now have the relaxed optimization problem
\begin{equation}
\begin{split}
    \mathrm{minimize} & \ \ \ \  \sum_{\mathbf{x}\in D} p(\mathbf{x}, \{f, \mathbf{W}+\Delta\mathbf{W}\}) \\
    \mathrm{s.t.}     & \ \ \ \  \mathcal{L}(\Delta\mathbf{W}) \leq th_g
\end{split}
\end{equation}

For this relaxed problem, we can utilize Lagrange multiplier to provide an alternative formulation
\begin{equation}
    \mathrm{minimize} \left( c \cdot \sum_{\mathbf{x}\in D} p(\mathbf{x}, \{f, \mathbf{W}+\Delta\mathbf{W}\}) + (\mathcal{L}(\Delta\mathbf{W}) - th_g)\right)\label{eq:final}
\end{equation}
where $c > 0$ is a suitably chosen constant, if optimal solution exists. This objective is equivalent to the relaxed problem, in the sense that there exists $c > 0$ such that the optimal solution to the latter matches the optimal solution to the former. 

Thus, we use the optimization objective (\ref{eq:final}) as the relaxed alternative objective of the defined objective (\ref{eq:ori_def}). Because the objective (\ref{eq:final}) is differentiable \emph{w.r.t.} $\Delta\mathbf{W}$, we use gradient descent as the optimization algorithm to solve this problem.

\textbf{The choice of constant $c$.}

\begin{figure}[ht]
    \vspace{-0.4cm}
    \includegraphics[trim=0 140 260 0, clip, width=0.8\linewidth]{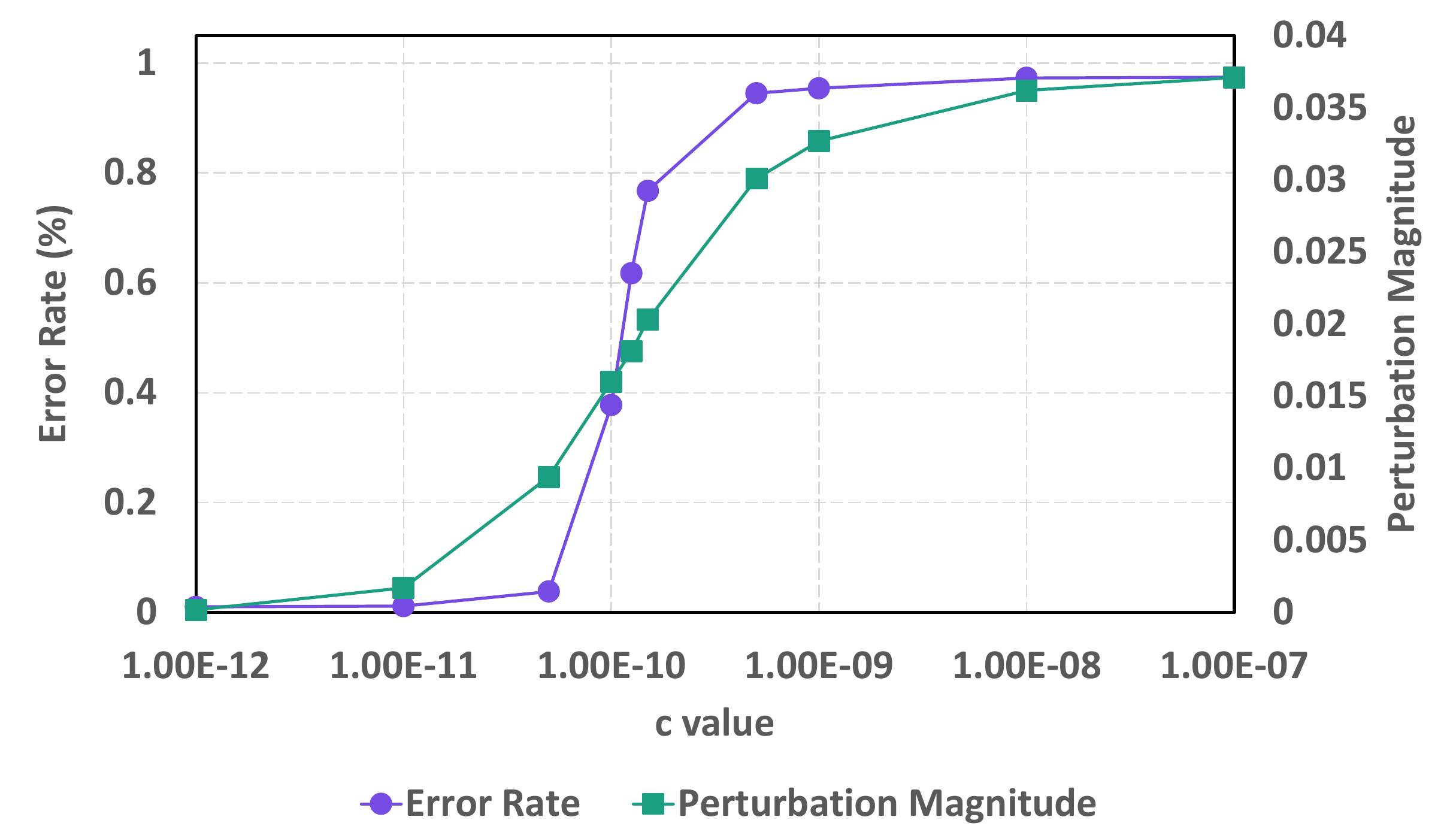}
\vspace{-0.4cm}
\caption{Choice of the constant $c$. We plot two different relations: (1) Worst-case error rate (left) \emph{w.r.t.} $c$ value and (2) perturbation magnitude $\mathcal{L}(\Delta\mathbf{W})$ (right) \emph{w.r.t.} $c$ value. The DNN model used is LeNet for MNIST.}
\vspace{-0.4cm}
\label{fig:CvA}
\end{figure}

Qualitatively speaking, observing objective (\ref{eq:final}) with larger $c$ values means more focus on lower accuracy and less focus on $\mathcal{L}(\Delta\mathbf{W})$, which would result in lower final accuracy and greater $\mathcal{L}(\Delta\mathbf{W})$. The empirical results shown in Fig.~\ref{fig:CvA} also prove this observation, where we plot how the worst-case error rate and $\mathcal{L}(\Delta\mathbf{W})$ varies with the choice of c using LeNet for MNIST. 

Because the empirical result show that $\mathcal{L}(\Delta\mathbf{W})$ is monotonic \emph{w.r.t.} $c$, to find the $c$ value that leads to the lowest performance under the weight perturbation bound $th_g$, we use binary search to find the largest $c$ value that ensures $\mathcal{L}(\Delta\mathbf{W}) \leq th_g$. The corresponding accuracy obtained with this $c$ is then the worst-case performance of this DNN model under weight perturbations bounded by $th_g$.

Finally, since the optimization problem stated in (\ref{eq:final}) can be solved via gradient descent, the time and memory complexity of our algorithm is comparable to that needed to train the DNN. 

%% file: M4_GPO_Experiments.tex
\begin{table}[ht]
    \centering
    \caption{Hyper-parameter setups to perform the proposed method on different models for different datasets. The $c$ value in (\ref{eq:final}) identified by binary search, the learning rate (lr) and the number of iterations used for gradient descent are specified.}
    \vspace{-0.3cm}
    \begin{tabular}{cccccc}
        \toprule
        Dataset         & Model     & c    & lr & \# of runs\\
        \midrule
        MNIST           & LeNet     & 1E-3  & 1E-5 & 500\\
        CIFAR-10        & ConvNet   & 1E-5  & 1E-5 & 100\\
        CIFAR-10        & ResNet-18 & 1E-9  & 1E-4 & 20 \\
        Tiny ImgNet     & ResNet-18 & 1E-10 & 1E-4 & 20 \\
        ImageNet        & ResNet-18 & 1E-3  & 1E-3 & 10 \\
        ImageNet        & VGG-16    & 1E-3  & 1E-3 & 10 \\
        \bottomrule
    \end{tabular}
    \label{tab:GPO_setup}
    \vspace{-0.4cm}
\end{table}
\subsection{Experimental Evaluation}\label{sect:exp}

In this section, we use experiments to demonstrate the effectiveness of the proposed method in finding the worst-case performance of different DNN models. Six different DNN models for four different datasets are used: (1) LeNet~\cite{lecun1998gradient} for MNIST~\cite{deng2012mnist}, (2) ConvNet~\cite{peng2019dnn+} for CIFAR-10~\cite{krizhevsky2009learning}, (3) ResNet-18~\cite{he2016deep} for CIFAR-10, (4) ResNet-18 for Tiny ImageNet~\cite{le2015tiny}, (5) ResNet-18 for ImageNet~\cite{deng2009imagenet} and (6) VGG-16~\cite{simonyan2014very} for ImageNet. LeNet and ConvNet are quantized to 4 bits while ResNet-18 and VGG-16 are quantized to 8 bits. As discussed in Section~\ref{sect:model}, we use $th_g = 0.03$ as the weight perturbation bound, \emph{i.e.}, each weight is perturbed by at most $\pm 0.03$. 

As there is no existing work on identifying the worst-case performance of a DNN under device variations to compare with other than the naive MC simulations, we slightly modify the weight PGD attack method~\cite{tsai2021formalizing}, which tries to find the smallest weight perturbation that can lead to a successful attack, as an additional baseline. Experiments are conducted on Titan-XP GPUs in the PyTorch framework. For MC simulation baseline, 100,000 runs are used. We use Adam~\cite{kingma2014adam} as the gradient descent optimizer. The detailed setup for the proposed method is shown in Table~\ref{tab:GPO_setup}.


\begin{table*}[t]
    \centering
    \caption{Comparison between MC simulation (MC), weight PGD attack (PGD) and the proposed framework in obtaining the worst-case accuracy of various DNN models for different dataset using weight perturbation bound $th_g = 0.03$. The accuracy of the original model without perturbation (Ori. Acc) is also provided. The proposed method finds perturbations that lead to much lower accuracy than those found by other methods, using slightly longer time than the weight PGD attack method but much shorter time than the MC simulation.}
    \vspace{-0.3cm}
    \begin{tabular}{cccrrrrrr}
        \toprule
        \multirow{2}{*}{Dataset} & \multirow{2}{*}{Model}          &\multirow{2}{*}{Ori. Acc.} & \multicolumn{3}{c}{Worst-case Accuracy (\%)} & \multicolumn{3}{c}{Time (Minutes)}\\ 
                        &     &    & MC & PGD & Proposed & MC & PGD & Proposed\\
        \midrule
        MNIST           & LeNet     & 99.12 & 97.34  & 13.44 & \textbf{7.35} & 900    & 3.3  & 5.5\\
        CIFAR-10        & ConvNet   & 85.31 & 60.12  & 10.00 & \textbf{4.27} & 2700   & 4.2  & 6.0\\
        CIFAR-10        & ResNet-18 & 95.14 & 88.77  & 10.00 & \textbf{0.00} & 5400   & 13.3 & 20.0\\
        Tiny ImageNet         & ResNet-18 & 65.23 & 25.33  & 0.50 & \textbf{0.00} & 14400  & 40.0 & 60.0\\
        ImageNet        & ResNet-18 & 69.75 & 43.98  & 0.10 & \textbf{0.00} & 231000 & 1980 & 2880\\
        ImageNet        & VGG-16    & 71.59 & 66.43  & 0.10 & \textbf{0.06} & 313800 & 2530 & 3820\\
        \bottomrule
    \end{tabular}
    \label{tab:GPO_res}
    \vspace{-0.3cm}
\end{table*}

\subsubsection{Worst-case DNN Accuracy Obtained by Different Methods}

As shown in Table~\ref{tab:GPO_res}, compared with weight PGD attack and MC simulations, the proposed framework is more effective in finding the worst-case performance. 
It identifies worst-case weight perturbations that can lead to below $10\%$ accuracy for LeNet and ConvNet, and almost $0\%$ accuracy for ResNet-18 and VGG-16.  On the other hand, the weight PGD attack can only find perturbations that lead to DNN accuracy close to random guessing (i.e., 1/N for N classes, which is 10\% for CIFAR-10, 0.5\% for Tiny ImageNet, and 0.1\% for ImageNet). MC simulations perform the worst. With 100,000 runs it fails to find any perturbation that can 
result in accuracy drop similar to 
those of the other two methods. This is quite expected given the high dimensional exploration space spanned by the large number of weights. Our framework takes slightly longer time to 
run than the weight PGD attack method, mainly due to the number of epochs the 
gradient descent takes to converge. Yet both methods are much faster than 
the MC simulations. 

The results from the table suggest that DNNs are extremely vulnerable to device variations, even though write-verify is used and the maximum weight perturbation is only $0.03$. Considering the fact that even converged 100,000 MC simulations cannot get close to the actual worst-case accuracy, For safety-critical applications, it may be necessary to screen each programmed CiM accelerator and test its accuracy to avoid disastrous consequences. Random sampling based quality control may not be an option. 

In addition, comparing the results obtained by our framework on ConvNet and ResNet-18 for CIFAR-10 (as well as ResNet-18 and VGG-16 for ImageNet) we can see that deeper networks are more susceptible to weight perturbations. This is expected as more perturbation can be accumulated in the forward propagation. 

Finally, the experimental results also show that quantization in both weights and activations is not an effective method to improve worst-case DNN performance, because all the models in these experiments are quantized as explained in the experimental setup.

\subsubsection{Analysis of Classification Results}

\ 

\begin{figure}[ht]
    \centering
    \vspace{-0.4cm}
    \includegraphics[trim=0 0 0 0, clip, width=0.8\linewidth]{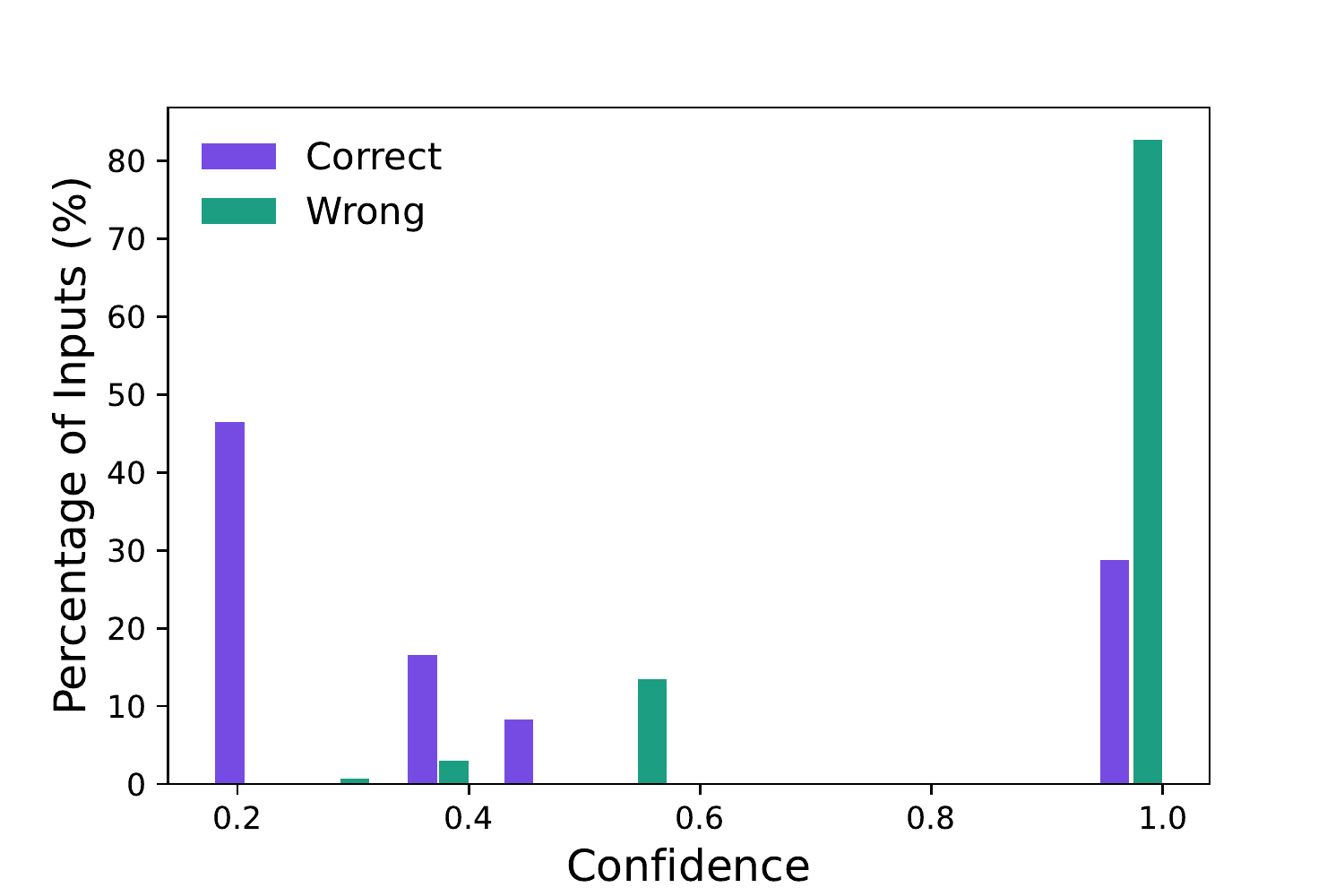}
    \vspace{-0.4cm}
    \caption{Distribution of classification confidence among correct/wrong cases from the worst-case LeNet for MNIST. The model is more confident in the wrong cases than in the correct ones. }
    \label{fig:conf_dist}
    \vspace{-0.4cm}
\end{figure}

We now take a closer look at the classification results of the worst-case LeNet for 
MNIST identified by our framework. We first examine the classification confidence, the distribution of which is shown in Fig.~\ref{fig:conf_dist}. Same as the common practice, the classification confidence of a DNN on an input is calculated by a Softmax function on its output vector. The element having the highest confidence is considered the classification result. Contrary to our intuition, from the figure we can see that the worst-case LeNet is highly confident in the inputs it mis-classifies, with an average confidence of 0.90 on all the inputs that are classified wrong. On the other hand, the DNN model is not confident in the inputs it classifies correctly, having an average confidence of only 0.47 on these inputs. This is significantly different from the original LeNet without perturbation, whose confidence is always close to 1.

We also observe how the classification results are distributed among different classes, which are reported in Table.~\ref{tab:class}.  From the table we can see that most of the 
errors are due to images being wrongly classified to the same class (class 1), while many 
of the images that truly belong to this class are being classified to other classes (class 2 and class 3). 

We hope that these observations can potentially shed light on the development of new algorithms to boost the worst-case performance of DNNs in the future. 


\begin{table}[ht]
    \centering
    \vspace{-0.2cm}
    \caption{Normalized classification results of the worst-case LeNet for MNIST. The number in row $i$ and column $j$ indicates how many cases with class $i$ as ground truth are being classified as $j$, normalized over the total number of cases in class $i$. Most inputs are mis-classified to one class (class 1).}
    \vspace{-0.3cm}
    \begin{tabular}{cc|cccccccccc}
        \toprule
        & &\multicolumn{10}{c}{Classification Result}\\
        & & 0 & 1 & 2 & 3 & 4 & 5 & 6 & 7 & 8 & 9 \\ 
        \midrule
        \multirow{10}{*}{\rotatebox[origin=c]{90}{Ground Truth}}
        &0& 0.0 & \red{1.0} & 0.0 & 0.0 & 0.0 & 0.0 & 0.0 & 0.0 & 0.0 & 0.0 \\ 
        &1& 0.0 & 0.4 & \red{0.5} & \red{0.1} & 0.0 & 0.0 & 0.0 & 0.0 & 0.0 & 0.0 \\ 
        &2& 0.0 & \red{1.0} & 0.0 & 0.0 & 0.0 & 0.0 & 0.0 & 0.0 & 0.0 & 0.0 \\ 
        &3& 0.0 & \red{1.0} & 0.0 & 0.0 & 0.0 & 0.0 & 0.0 & 0.0 & 0.0 & 0.0 \\ 
        &4& 0.0 & \red{0.9} & 0.0 & 0.0 & 0.0 & 0.0 & 0.0 & \red{0.1} & 0.0 & 0.0 \\ 
        &5& 0.0 & \red{0.9} & 0.0 & \red{0.1} & 0.0 & 0.0 & 0.0 & 0.0 & 0.0 & 0.0 \\ 
        &6& 0.0 & \red{1.0} & 0.0 & 0.0 & 0.0 & 0.0 & 0.0 & 0.0 & 0.0 & 0.0 \\ 
        &7& 0.0 & \red{0.6} & 0.0 & \red{0.4} & 0.0 & 0.0 & 0.0 & 0.0 & 0.0 & 0.0 \\ 
        &8& 0.0 & \red{0.8} & 0.0 & \red{0.2} & 0.0 & 0.0 & 0.0 & 0.0 & 0.0 & 0.0 \\ 
        &9& 0.0 & \red{0.9} & 0.0 & \red{0.1} & 0.0 & 0.0 & 0.0 & 0.0 & 0.0 & 0.0 \\ 
        \bottomrule
    \end{tabular}
    \label{tab:class}
    \vspace{-0.4cm}
\end{table}

\subsubsection{Distribution of Worst-Case Weight Perturbation}

\ 

\begin{figure}[ht]
    \centering
    \vspace{-0.6cm}
    \includegraphics[trim=0 0 0 0, clip, width=0.8\linewidth]{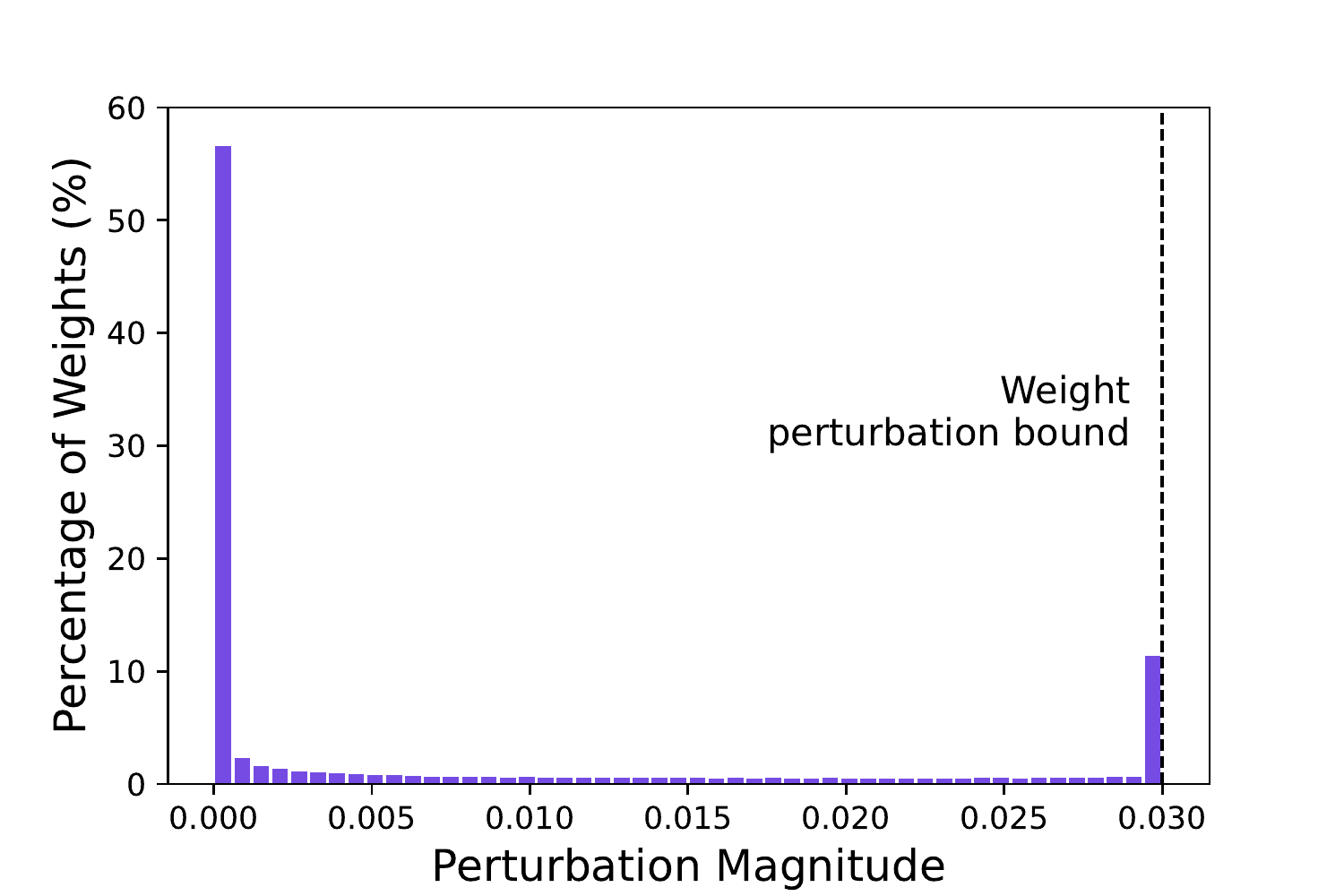}
    \vspace{-0.4cm}
    \caption{The distribution of the weight perturbation magnitude in the worst-case LeNet for MNIST. Most weights are either not perturbed or perturbed to $th_g$.}
    \label{fig:mag_dist}
\end{figure}

Here we show how the perturbation is distributed among the weights 
in the worst-case LeNet for MNIST. As can be seen in Fig.~\ref{fig:mag_dist}, 
most of the weights are either not perturbed or perturbed to the maximum magnitude (\emph{i.e.}, $th_g=0.03$).  

\begin{figure}[ht]
    \centering
    \vspace{-0.4cm}
    \includegraphics[trim=0 110 300 0, clip, width=0.7\linewidth]{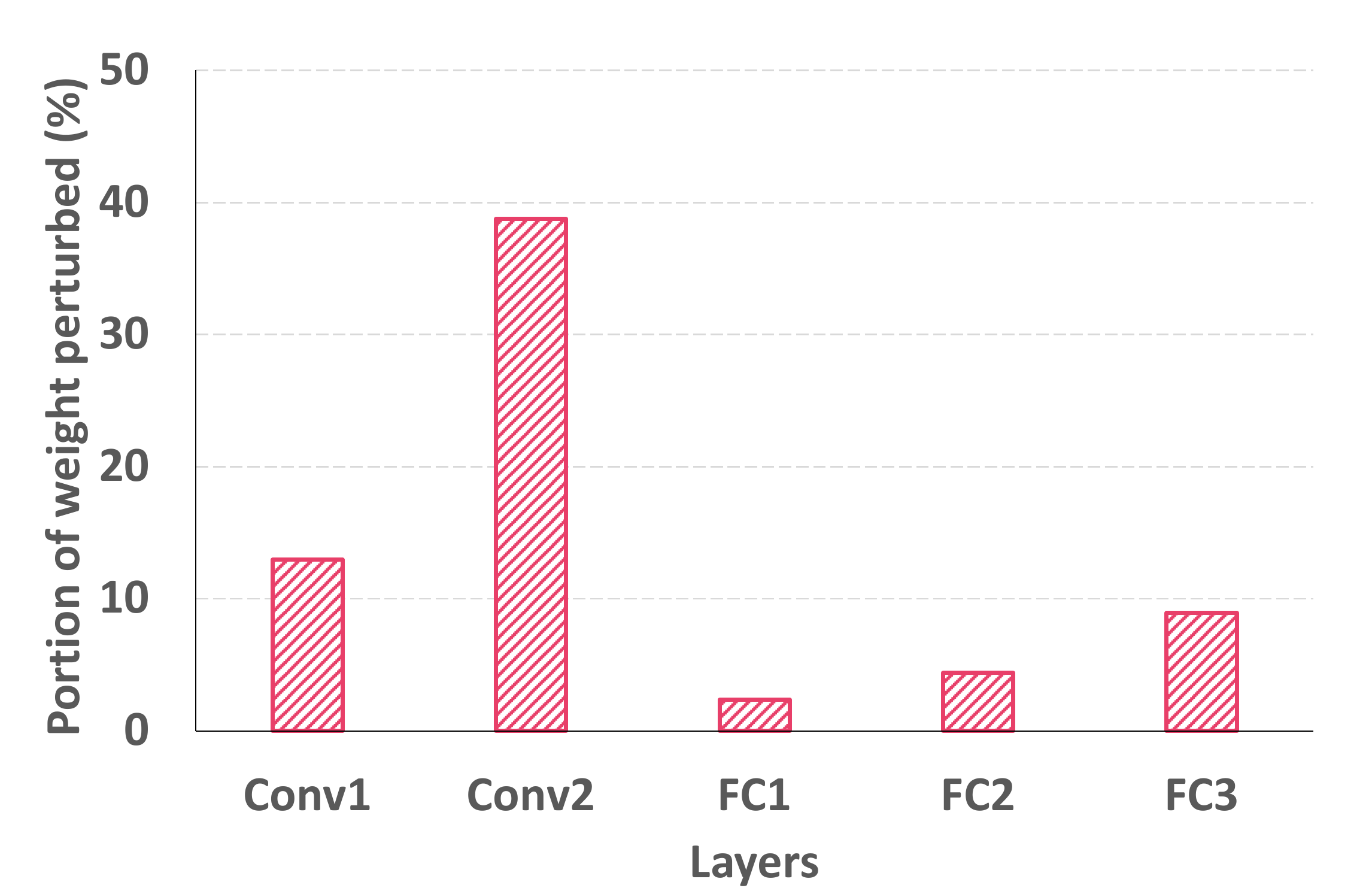}
    \vspace{-0.4cm}
    \caption{The percentage of weights being perturbed in each layer of the worst-case LeNet for MNIST. Weights in convolutional layers and the last FC layer are more likely to be perturbed.}
    \label{fig:layer}
    \vspace{-0.6cm}
\end{figure}

We then show the number of weights that are perturbed in each layer in Fig.~\ref{fig:layer}. We can observe that the weights in convolutional layers and the final FC layer are more likely to be perturbed. This is probably due to the fact that 
they in general have more impact on the accuracy of a DNN. 

%% file: M5_Protect.tex
\begin{figure*}
\vspace{-0.4cm}
\begin{minipage}[b]{0.32\linewidth}
    \includegraphics[trim=10 0 30 30, clip, width=\linewidth]{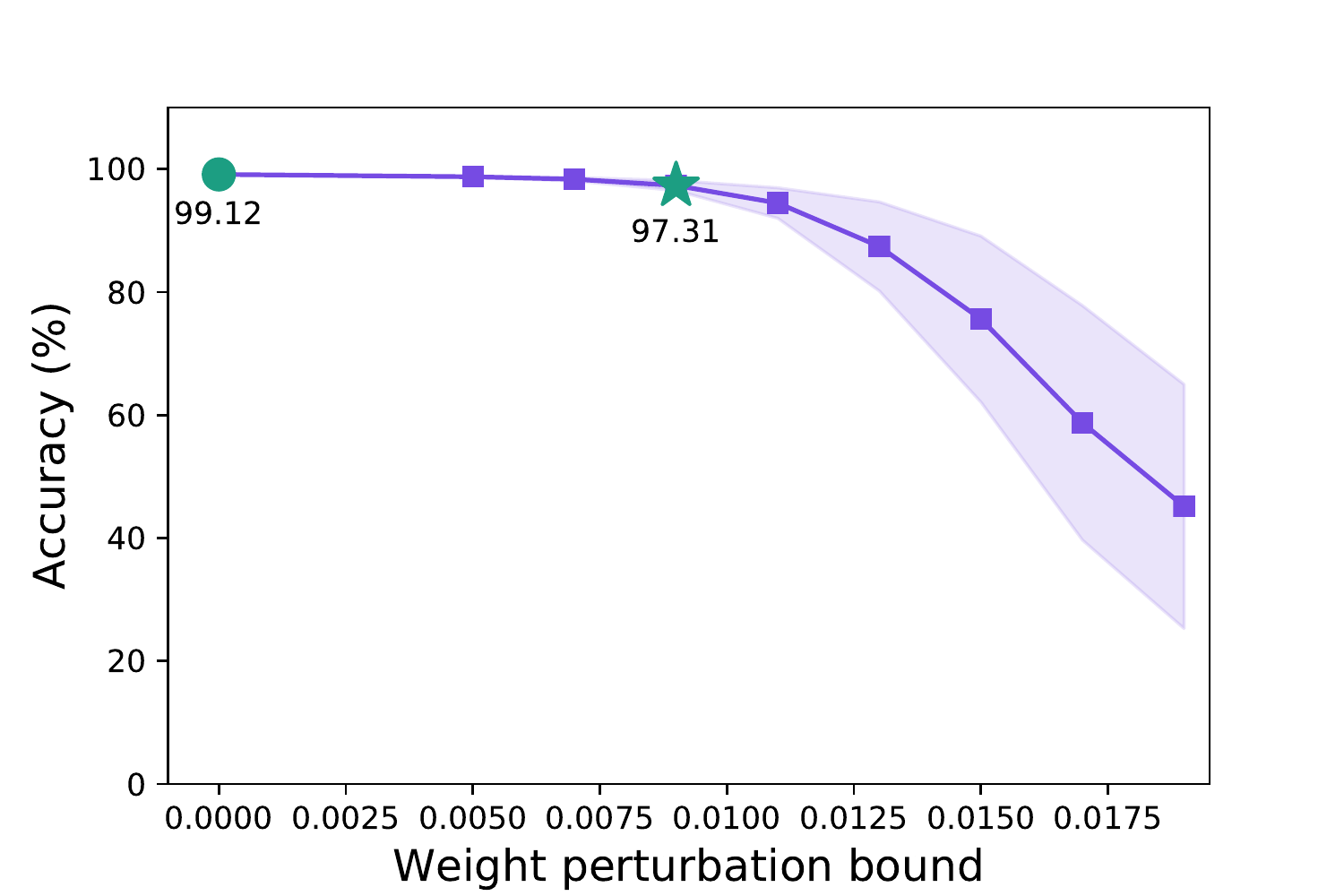}
    \vspace{-0.5cm}
    \subcaption{LeNet for MNIST \textbf{w/} reg. training.}
    \label{fig:diff_dist_lenet}
\end{minipage}
\begin{minipage}[b]{0.32\linewidth}
    \includegraphics[trim=8 0 32 30, clip, width=\linewidth]{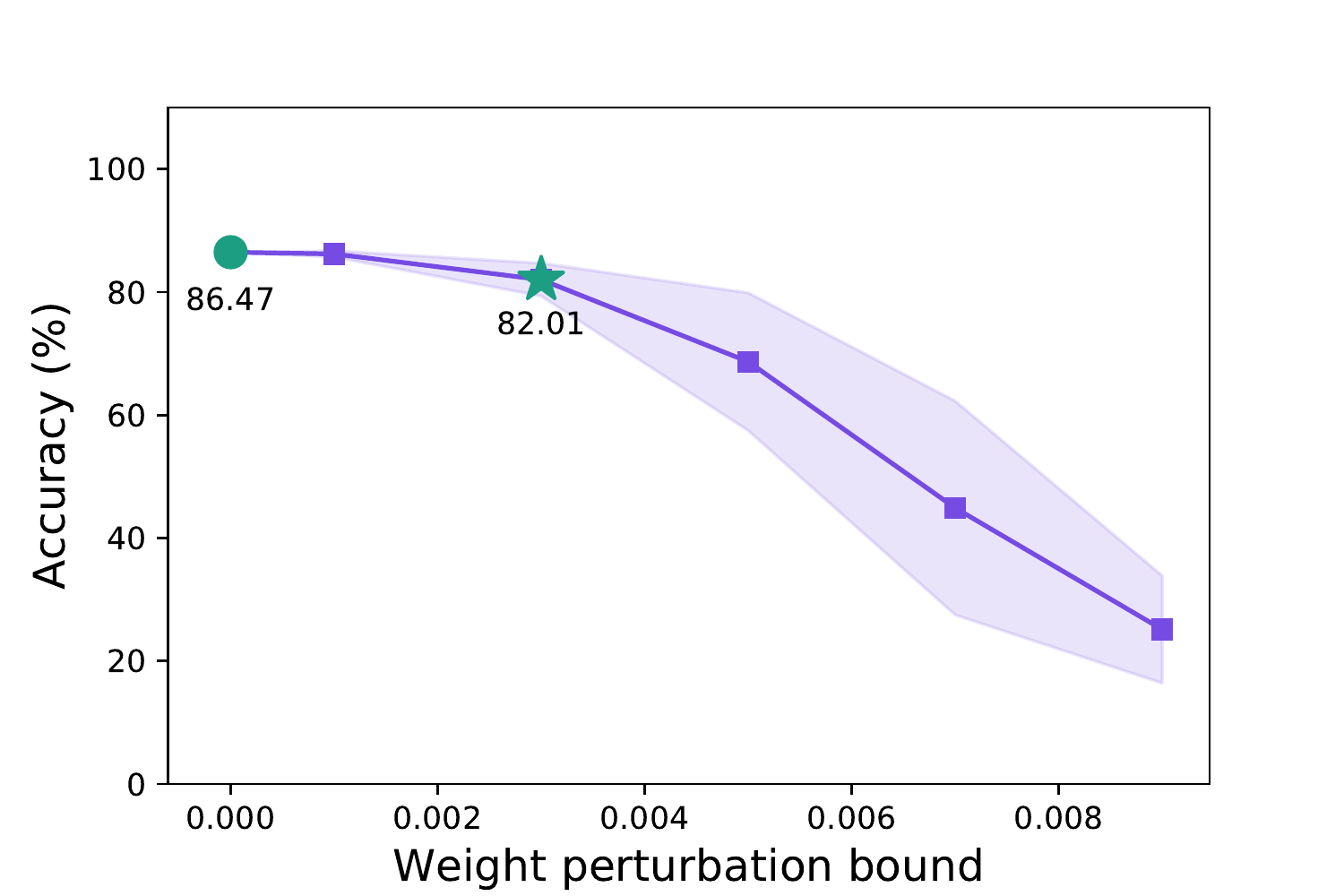}    
    \vspace{-0.5cm}
    \subcaption{ConvNet for CIFAR-10 \textbf{w/} reg. training.}
    \label{fig:diff_dist_cifar}
\end{minipage}
\begin{minipage}[b]{0.32\linewidth}
    \includegraphics[trim=10 0 30 30, clip, width=\linewidth]{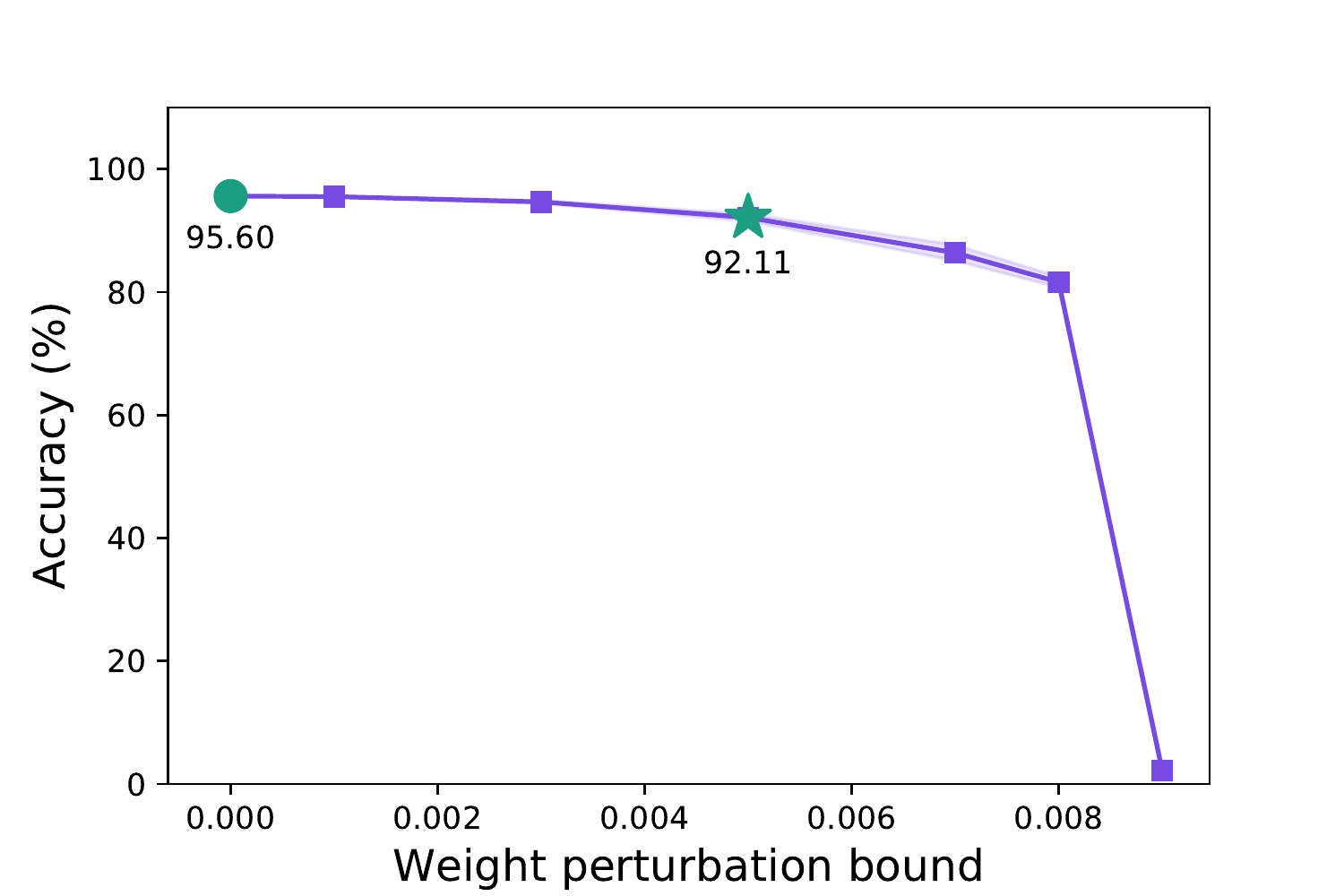}
    \vspace{-0.5cm}
    \subcaption{ResNet18 for CIFAR-10 \textbf{w/} reg. training.}
    \label{fig:diff_dist_res18}
\end{minipage}
\begin{minipage}[b]{0.32\linewidth}
    \includegraphics[trim=10 0 30 30, clip, width=\linewidth]{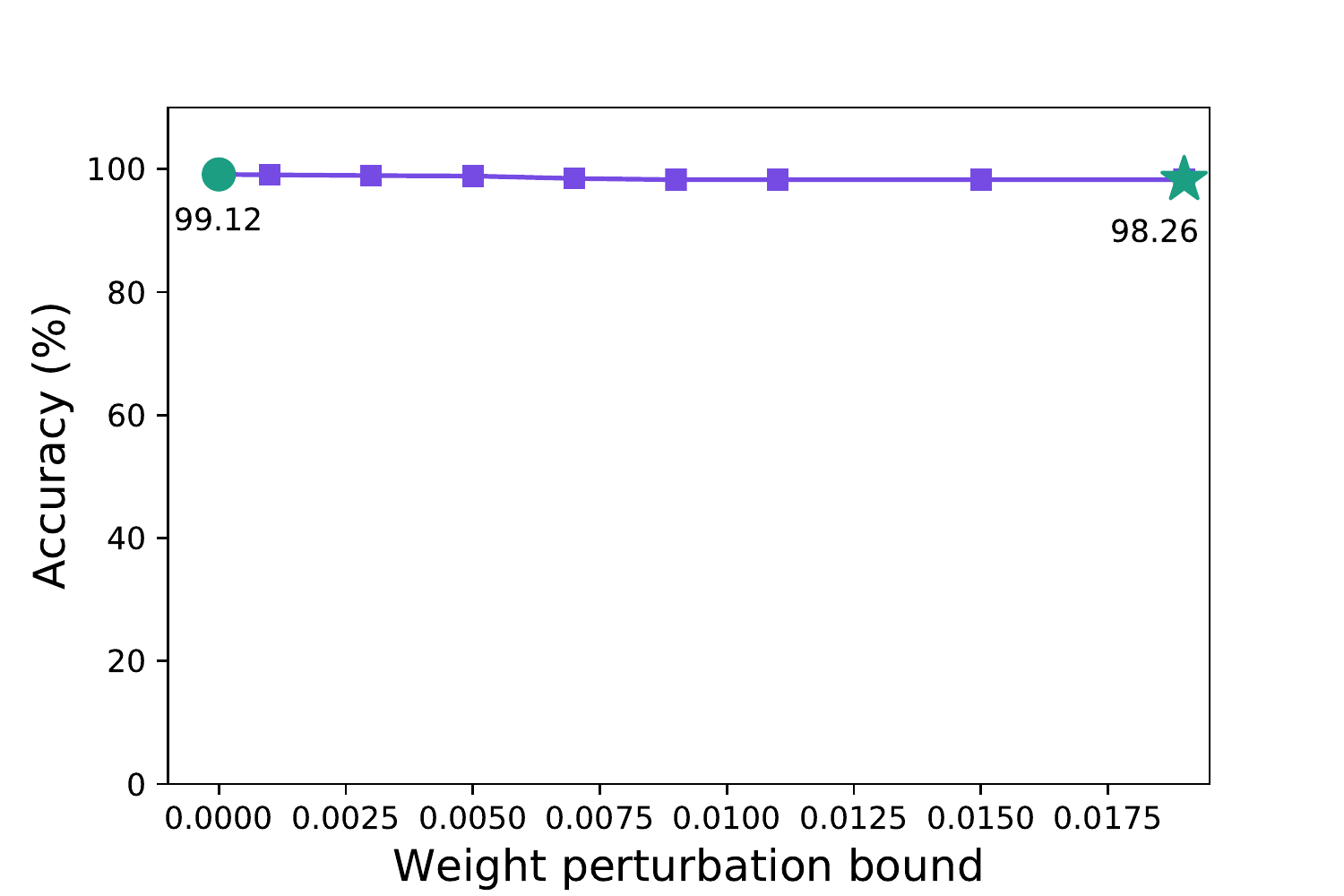}
    \vspace{-0.5cm}
    \subcaption{LeNet for MNIST \textbf{w/} adv. training.}
    \label{fig:diff_AN_lenet}
\end{minipage}
\begin{minipage}[b]{0.32\linewidth}
    \includegraphics[trim=8 0 32 30, clip, width=\linewidth]{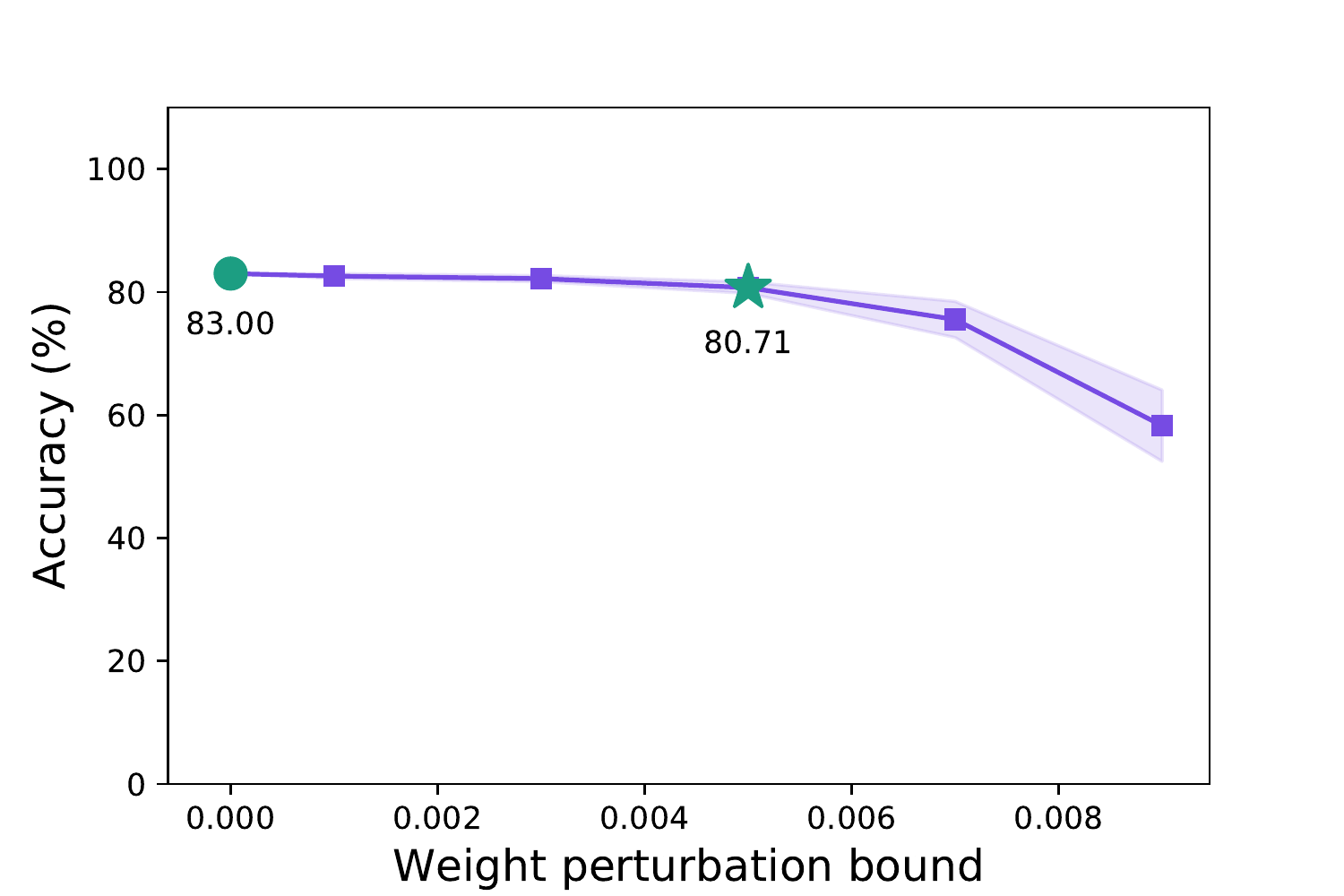}    
    \vspace{-0.5cm}
    \subcaption{ConvNet for CIFAR-10 \textbf{w/} adv. training.}
    \label{fig:diff_AN_cifar}
\end{minipage}
\begin{minipage}[b]{0.32\linewidth}
    \includegraphics[trim=10 0 30 30, clip, width=\linewidth]{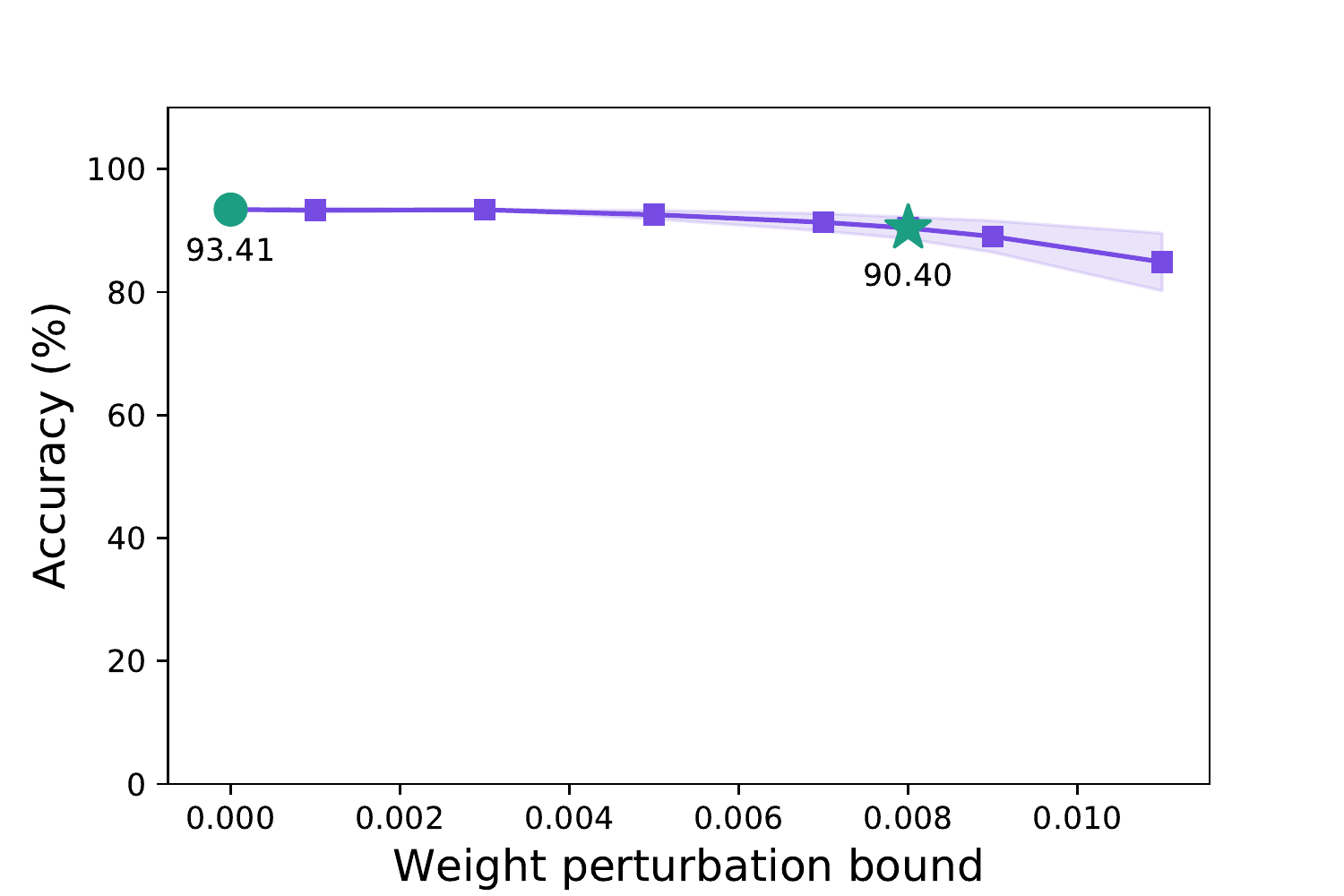}
    \vspace{-0.5cm}
    \subcaption{ResNet-18 for CIFAR-10 \textbf{w/} adv. training.}
    \label{fig:diff_AN_res18}
\end{minipage}
\vspace{-0.3cm}

\caption{Effectiveness of write-verify with regular training (a)-(c), and with adversarial training (d)-(f). Figures represent the relationship between weight perturbation bound 
in write-verify $th_g$ (X-axis) and the worst-case DNN accuracy (Y-axis) in different models: (a)(e) LeNet for MNIST, (b)(d) ConvNet for CIFAR-10, and (c)(f) ResNet-18 for CIFAR-10. For each data point, three experiments of the same setting but different random initialization are conducted. The solid lines show the averaged results over the three experiments and the shadows represent the standard deviations. In each figure, the circle marks the model without perturbation ($th_g=0$) and the star marks the model with highest $th_g$ and no more than $5\%$ accuracy degradation.}
\vspace{-0.5cm}
\label{fig:all_ORI}
\end{figure*}

\section{Enhancing Worst-Case Performance of CiM DNN Accelerators}\label{sect:protect}
Several works exist in the literature to improve the average performance of a DNN
under device variations. In this section, we try to extend them to improve the 
worst-case performance of a DNN, and evaluate their effectiveness. 
Specifically, we will include two types of methods: (1) confining the device variations and (2) training DNN models that are more robust against device variations. As discussed in Section~\ref{sect:related}, one of the most popular practices of the former is write-verify and the latter includes variation-aware training.


In addition to these, 
we also modify adversarial training~\cite{tsai2021formalizing}, a method commonly used to combat
adversarial input, to address weight perturbation caused by device variations. 
The algorithm is summarized in Alg.~\ref{alg:adv}. Similar to how adversarial training handles input perturbations, in DNN training process we inject worst-case perturbations to the weights of a DNN, in hope that they perform better under the impact of device variations. Specifically, in each iteration of training, we first conduct the proposed method to find the perturbations of the current weights tha can lead to worst-case accuracy of the model. We then add them to the weights and collect the gradient $G$. 

If not specified explicitly, all accuracy results shown in this section are collected by training one DNN architecture using the same specification but with \textbf{three} different initializations. The accuracy (error rate) number is shown in percentage and presented in [average $\pm$ standard deviation] for these three runs.

\begin{algorithm}[ht]
\caption{Adversarial Training~($f$,  $D$, $\mathbf{V}$, $Ep$, $l$, $\eta$, $\mathbf{C}$)}
\begin{algorithmic}[1]\label{alg:adv}
\STATE // INPUT: A DNN architecture $f$, training dataset $\mathbf{D}$, validation dataset $\mathbf{V}$, the total number of training epochs $Ep$, loss function $l$, learning rate $\eta$;

\STATE Initialize weight $\mathbf{W}$ for $f$;
\STATE Initialize $acc_F = 0$, $\mathbf{W}_F = \mathbf{W}$;

\FOR{($i=0$; $i < Ep$; $i++$)}

    \FOR{mini-batches $\mathbf{B}$ in $\mathbf{D}$}
        \STATE Divide $\mathbf{B}$ into input $\mathbf{I}$ and label $\mathbf{L}$;
        \STATE Find weight perturbations $\mathbf{N}$ that lead to worst-case accuracy using the framework discussed in Section~\ref{sect:method};
        \STATE $\mathbf{O} = f(\mathbf{W} + \mathbf{N}, \mathbf{I})$;
        \STATE $loss = l(\mathbf{O}, \mathbf{L})$;
        \STATE $\mathbf{G} = \frac{\partial loss}{\partial \mathbf{W}}$;
        \STATE $\mathbf{W} = \mathbf{W} - \eta \times \mathbf{G}$
        
    \ENDFOR
    \STATE Evaluate $\mathbf{W}$ on $\mathbf{V}$ and get $acc$;
    \IF{$acc > acc_F$}
        \STATE $acc_F = acc$;
        \STATE $\mathbf{W_F} = \mathbf{W}$
    \ENDIF
\ENDFOR
\STATE Return $\mathbf{W}_F$
\end{algorithmic}
\end{algorithm}

\subsection{Stronger Write-Verify}\label{sect:wv}

As shown in Table~\ref{tab:GPO_res}, using the standard write-verify setting in the literature to confine the maximum weight perturbation to 0.03 ($th_g = 0.03$ in (\ref{eq:noise})) cannot significantly improve the worst-case performance of DNN models. If we set a smaller $th_g$ in write-verify, the write time would become longer but can potentially help to boost the worst-case performance. 

To see the relationship between $th_g$ and worst-case DNN accuracy, we use three models, \emph{i.e.}, LeNet for MNIST, ConvNet for CIFAR-10, and ResNet-18 for CIFAR-10, and plot the results 
as shown in Fig.~\ref{fig:all_ORI}(a)-(c), where we also include the model accuracy without any device variation ($th_g=0$). From the figures we can see that a lower $th_g$ can indeed increase the worst-case accuracy. Yet to ensure the models have acceptable accuracy in worst-case (\emph{e.g.}, no more than 5\% accuracy drop from DNNs without the impact of device variations and marked with star in each figure), we need to set $th_g = 0.009$ for LeNet, $th_g = 0.003$ for ConvNet and $th_g = 0.005$ for ResNet-18, which would take extremely long write time. The experimental result of ResNet-18 for Tiny ImageNet is not shown here because its worst-case accuracy is lower than 20\% even when $th_g = 0.001$ and further reducing $th_g$ is not practical.

\subsection{Variation-aware and Adversarial Training}\label{sect:tm}

 \begin{table}[t]
    \centering
    \vspace{-0.4cm}
    \caption{Worst-case accuracy (\%) of various DNN models from regular training (Regular), variation-aware training (VA) and adversarial training (ADV). Write-verify with weight perturbation bound $th_g = 0.03$. Compared with regular training, adversarial training is effective in LeNet for MNIST, but both methods are not effective in other more complex models.}
    \vspace{-0.4cm}
    \begin{tabular}{ccccc}
        \toprule
        Dataset         & Model     & Regular    & VA & ADV \\
        \midrule
        MNIST           & LeNet     & 7.35$\pm$03.70 & 18.58$\pm$00.80 & 98.26$\pm$01.05\\
        CIFAR10         & ConvNet   & 4.27$\pm$00.33 & 63.71$\pm$03.76 & 67.09$\pm$03.85\\
        CIFAR10         & ResNet18  & 0.00$\pm$00.00 & 32.84$\pm$17.20 & 34.84$\pm$13.20\\
        Tiny IN         & ResNet18  & 0.00$\pm$00.00 & 3.57$\pm$03.48 & 7.41$\pm$08.10\\
        \bottomrule
    \end{tabular}
    \label{tab:protect}
    \vspace{-0.4cm}
\end{table}

Here we study the effectiveness of variation-aware training and adversarial training on four models: LeNet for MNIST, ConvNet for CIFAR-10, ResNet-18 for CIFAR-10, and ResNet-18 for Tiny ImageNet. We assume that standard write-verify with weight perturbation bound $th_g=003$ is used for all the models. 

As shown in Table~\ref{tab:protect}, both variation-aware training and adversarial training can offer some improvements in most cases compared with the regular training. Adversarial training is slightly more effective than variation-aware training. However, compared with the accuracy that can be obtained by these networks without device variations (third column in Table~\ref{tab:GPO_res}), the accuracy drop 
is still significant in the worst case. The only exception is the case of LeNet for MNIST, where adversarial training can almost fully recover the accuracy loss even in the worst case, thanks to its simplicity. In addition, we can observe that as the network gets deeper, the worst-case accuracy improvement brought by these two training methods starts to diminish (e.g. $7.41\%$ for ResNet-18 for Tiny ImageNet). 

\subsection{Combining Adversarial Training with Write-Verify}

Finally, using the same three models and datasets, we show whether the models trained by the adversarial training method can reduce the requirement on write-verify to achieve the same worst-case accuracy. The results are shown in Fig.~\ref{fig:all_ORI}(d)-(f). 
Comparing with the results of the models from regular training in Fig.~\ref{fig:all_ORI}(a)-(c), with adversarial training, the weight perturbation bound $th_g$ needed to achieve the same accuracy increases. As discussed in Section~\ref{sect:tm}, with adversarial training, the worst-case accuracy of LeNet for MNIST using the standard write-verify ($th_g = 0.03$) is already very close to that of the original model without device variations. Thus, Fig.~\ref{fig:all_ORI}~(d) is almost flat. For the other two models, to ensure no more than 5\% worst-case accuracy degradation from the original model without device variations, we now need $th_g = 0.005$ for ConvNet for CIFAR-10, and $th_g = 0.008$ for ResNet-18 for CIFAR-10, as marked by the star in each figure.
Comparing with the weight perturbation bound needed to attain the same worst-case accuracy in Fig~\ref{fig:all_ORI}(b)-(c), we can see that using adversarial training instead of regular training can increase it by around 1.7$\times$, indicating faster programming process. However, these bounds are still much smaller than the commonly used $0.03$~\cite{jiang2020device, shim2020two} and take considerably more programming time. 
Therefore, more effective methods to address 
the worst-case accuracy are still needed, and calls for future research. 

%% file: M6_Conclusions.tex
\section{Conclusions}\label{sect:conclusion}
In this work, contrary to the existing methods that evaluate the average performance of DNNs under device variations in CiM accelerators, we proposed an efficient framework to examine their worst-case performance, which is important for safety-critical applications. With the proposed framework, we show that even with bounded small weight perturbations after write-verify, the accuracy of a well-trained DNN can drop drastically to almost zero. As such, we should use caution when applying CiM accelerators to safety-critical applications. For example, we may need to screen the accuracy of each chip rather than random sampling in quality control.  We further show that the existing methods used to enhance average DNN performance in CiM accelerators are either too costly (for stronger write-verify) or ineffective (for training-based methods) when extended to enhance the worst-case performance. Further research from the community is needed to address this problem. 